\documentclass[12pt]{article}
\usepackage[english,activeacute]{babel}
\usepackage{natbib}
\usepackage{comment}
\usepackage{float}
\usepackage[hidelinks]{hyperref}
\usepackage{mathrsfs}
\usepackage{enumitem}
\usepackage[font={small,it}]{caption}
\usepackage{amsmath,amsfonts,amsthm,amssymb}
\usepackage{bm,rotating,multirow,dsfont,graphicx}
\usepackage[usenames,dvipsnames]{color}
\usepackage{url}
\usepackage{multicol}
\usepackage{multirow}
\usepackage[T1]{fontenc}
\usepackage{flafter}
\usepackage{appendix}
\usepackage{subfigure}
\usepackage{xcolor}
\usepackage{soul}
\usepackage{threeparttable}
\makeatletter
\def\hlinewd#1{%
	\noalign{\ifnum0=`}\fi\hrule \@height #1 %
	\futurelet\reserved@a\@xhline}
\makeatother
\newcommand{\be}{\begin{eqnarray*}}
\newcommand{\ee}{\end{eqnarray*}}
\newcommand{\bet}{\begin{eqnarray}}
\newcommand{\eet}{\end{eqnarray}}
\def\spacingset#1{\renewcommand{\baselinestretch}{#1}\small\normalsize}\spacingset{1}
\def\@roman#1{\romannumeral #1}
\graphicspath{{./plots/}}

\begin{document}

\title{Analysis of Bipartite Networks in Anime Series: \\ Textual Analysis, Topic Clustering, and Modeling}

\author{
    Juan Sosa$^{1}$\footnote{Email: jcsosam@unal.edu.co.} \\
    Alejandro Urrego-Lopez$^{1}$\footnote{Email: aurrego@unal.edu.co.} \\
    Cesar Prieto $^{1}$\footnote{Email: ceprieto@unal.edu.co.}
}

\date{
    $^{1}$Universidad Nacional de Colombia, Bogotá, Colombia
}

\maketitle


\begin{abstract} 
\noindent
This article analyzes a specific bipartite network that shows the relationships between users and anime, examining how the descriptions of anime influence the formation of user communities. In particular, we introduce a new variable that quantifies the frequency with which words from a description appear in specific word clusters. These clusters are generated from a bigram analysis derived from all descriptions in the database. This approach fully characterizes the dynamics of these communities and shows how textual content affect the cohesion and structure of the social network among anime enthusiasts. Our findings suggest that there may be significant implications for the design of recommendation systems and the enhancement of user experience on anime platforms.
\end{abstract}

\noindent
{\it Bigrams, Bipartite Networks, Graphs, Natural Language Processing, Network Partitioning, Exponential Random Graph Models.}

\spacingset{1.1} 

\section{Introduction}\label{intro}

The analysis of bipartite networks is a valuable method for studying the relationships between two sets of individuals. 
For instance, this approach can be utilized on websites that reflect users' interests in books or music, as well as on entertainment platforms where there is a connection between users and the movies they have viewed. 
Furthermore, we can examine the relationships between actors and the films they have participated in, as outlined in studies on bipartite networks and their application in modeling complex and collaborative systems \citep{wasserman1994}.

This approach is particularly useful for platforms that facilitate interactions between users and products, enabling the exploration of these relationships using statistical models.
One example involves modeling the choices made by customers on a specific platform or in a particular market. 
For instance, a study conducted in \cite{ghose2011} reexamined the impact of reviews on economic outcomes, such as product sales, and explored textual factors like subjectivity, readability, and spelling errors in the reviews to identify important textual characteristics.

Descriptions represent a relevant source of information for all platforms concerning product consumption, offering valuable insights into the features and attributes of the items available.
Composed of carefully selected words, such descriptions can be subjected to rigorous analysis using several text analysis techniques. 
By employing various statistical methods, researchers can identify distinctive patterns and trends emerging from the vocabulary and phrasing utilized in these descriptions. 
This analysis not only enhances the understanding of consumer perceptions but also facilitates the identification of key factors that influence purchasing decisions and overall user engagement \citep{manning2008}.

Descriptions, like any other nodal variable—referring to attributes possessed by individuals within a network—can influence the structural features of the network to which they are embedded.
Research in social network analysis shows how nodal characteristics shape network structures and influence connection patterns \citep{jackson2008}.
To assess the influence of a nodal variable within a network, one effective model is the Exponential Random Graph Model (ERGM; e.g., \citealt{frank1986markov}, \citealt{wasserman1996logit}, \citealt{robins2007introduction}), which allows for inferences about these nodal variables \citep{hunter2008}. 
Here, we analyze a dataset containing 16,214 anime series along with the preferences of 325,000 users, intentionally excluding those featuring adult content. Of these 16,214 records, 15,497 include complete descriptions. This publicly available data can be accessed at \url{https://github.com/Hernan4444/MyAnimeList-Database} and \url{https://www.kaggle.com/datasets/hernan4444/anime-recommendation-database-2020}.

The primary objective of this article, in addition to addressing several technical challenges and computational issues, is to evaluate the influence of anime series descriptions on the formation of large user communities surrounding these series.
This is particularly important as it helps us understand how anime series descriptions shape the formation of large user communities, providing valuable insights into crafting descriptions that effectively promote a series.
Moreover, it is evident that this approach can be applied to any platform utilizing descriptions in a similar way. Understanding how descriptions impact user engagement and community formation enables platforms to optimize their content strategies. For instance, social media and streaming services can create more engaging descriptions that inform users and foster community interaction, ultimately enhancing user retention and experience

This article is structured as follows: Section 2 presents the state of the art and relevant instances related to our research. Section 3 outlines the methodology and provides technical details necessary for conducting our analyses. Section 4 details the case study. Finally, Section 5 presents our main findings along with suggestions for future research directions.

\section{Literature Review}

In the context of recommendation systems, in this article we analyze bipartite networks as essential tools for modeling the interaction between two sets of nodes, such as users and products. 
This approach gains traction due to its effectiveness in representing the complex relationships that arise within recommendation systems. Studies like those of \cite{zhou2007} introduce a recommendation algorithm based on Random Walks on bipartite networks, which significantly enhances the accuracy of recommendations by incorporating greater diversity in product suggestions. Similarly, \cite{lu2012} propose a unified approach to the analysis of bipartite networks, demonstrating that these systems not only improve accuracy but also optimize the diversity of recommendations, a critical aspect for preventing the oversaturation of certain products.

Additionally, \cite{zhang2010} introduce an integrated diffusion method in bipartite networks, significantly improving the personalization of recommendations. This method effectively considers the connections between several nodes within the network, capturing the intricate relationships that exist between users and products. By leveraging these connections, the ability to tailor recommendations to individual user preferences is enhanced. Furthermore, these studies underscore the importance of content platforms in the recommendation process. They demonstrate that utilizing advanced techniques not only increases the accuracy of recommendations but also enhances the overall quality and diversity of the suggestions provided. This is crucial for ensuring that users receive a wide range of relevant options, ultimately leading to a more satisfying and engaging user experience.

On the other hand, regarding text analysis, the use of networks has proven to be a powerful methodology for studying the relationships between words and actors in various contexts. For instance, \cite{Luque2024} employ this methodology to characterize presidential speeches during specific presidential periods in Colombia. Similarly, \cite{gruhl2004} analyze the diffusion of information in blogs by modeling the relationship between texts and individuals as a network, which enables a deeper understanding of how information propagates within virtual communities. This technique has also been utilized to extract social networks from literary texts, as evidenced by the work of \cite{elson2010}, where natural language processing is applied to model social interactions based on written narratives.

Additionally, \cite{carley1992} apply textual network analysis to investigate organizational dynamics, demonstrating that word co-occurrence networks can provide valuable insights into the internal structure of organizations. By examining how words appear together, this approach uncovers relationships and interactions within the organization, revealing communication flows and decision-making processes. Moreover, such techniques are useful for understanding how descriptions of products, like anime series, influence the formation of communities on digital platforms. By identifying recurring patterns in these texts, researchers can uncover how language shapes user perceptions and engagement. This understanding can inform marketing strategies and content development, ultimately enhancing user experience and fostering stronger online communities.

Finally, exponential random graph models (ERGMs) have been crucial for modeling complex networks and extracting inferences about structural patterns. \cite{robins2007} review the theory and applications of ERGMs in social networks, highlighting their ability to capture phenomena such as transitivity and homophily. This approach is relevant for analyzing how node characteristics, such as series descriptions, influence network structure and the formation of user communities. Furthermore, the studies by \cite{frank1986} establish the theoretical foundations for Markovian modeling in networks, which would later be applied to fit real networks and model complex structures, such as user networks on entertainment platforms.

By integrating these approaches, we provide a novel methodology that not only improves the accuracy of recommendations on content platforms but also helps understand the influence of textual descriptions in the creation of communities, such as how product descriptions impact the structure and cohesion of user-formed networks. This perspective enables the development of an innovative technology that combines bipartite network analysis, social network extraction from text, and exponential random graph modeling, offering a deeper understanding of interactions on digital platforms.

The optimization of product descriptions and appeal in contemporary digital culture can have a direct impact on how fans perceive and engage with series. This phenomenon is understood as part of cultural creation, where products are consumed not only for their technical features but also for the symbolic and experiential value they convey. Recent studies suggest that descriptions capturing both symbolic and experiential value emotionally connect with users, enhancing consumer immersion and satisfaction (e.g., \citealt{qin2020metaphorical} and \citealt{plos2021}). However, to the best of our knowledge, there is currently no literature available on the use of quantitative methods specifically for optimizing descriptions in cultural products, which highlights the need for further research in this area.

\section{Methodology}

\subsection{Simple Random Sampling}

Simple Random Sampling (SRS; e.g., \citealt{groves2009}, \citealt{thompson2012} and \cite{kish2019}) is a key sampling technique in statistical research that ensures each element of a population has an equal chance of being selected. By randomly selecting a subset from the entire population, SRS minimizes bias and enhances the reliability of results, allowing researchers to generalize findings to the larger group. This method is particularly valuable in very field (such as this one!) where accurate population representation is crucial. SRS not only facilitates comparisons among different groups but also helps control for confounding variables, making it essential for sound statistical analysis and meaningful insights \citep{acharya2013sampling}.

Here, the target population consists of the anime series to be analyzed, defined as those that meet the criteria established in the study, including their availability on content platforms and the ratings received from users.
\cite{Krejcie1970} elaborate on the calculation of sample size for finite populations. When conducting SRS in a finite population, two common approaches exist for determining the sample size: One based on the proportion of a categorical characteristic of the population and the other based on the variance of a quantitative characteristic. In this case, we use the following expression to determine the sample size \(n\) based on the variance of the variable \textit{Score}, which represents the average rating given by users to the anime series:
\[
n = \frac{Z^{2} \cdot S^2 \cdot N}{(N-1) \cdot E^{2} + Z^{2} \cdot S^2}\,,
\]
where \(N\) is the total population size, \(Z\) is the critical value from the standard normal distribution that corresponds to the desired confidence level, \(S^2\) is the variance of the variable \textit{Score} within the population, and \(E\) is the permissible margin of error in estimating the mean.

To sample the database, anime classified as \textit{Hentai} is excluded because it is typically not available on the same platforms as other genres. Additionally, only genres with more than 100 occurrences are considered. This approach does not exclude anime from these genres, as a single anime may belong to multiple genres simultaneously; rather, it omits the categorization of genres with fewer than 100 occurrences.
For each genre, sampling is carried out based on the total population size, adjusting the confidence parameters and margin of error to achieve the desired level of precision. In this instance, we employ a confidence level of \(1 - \alpha = 0.85\) and a margin of error of \(\epsilon = 0.2\), resulting in a total of 1,314 anime for analysis.
However, because an anime can belong to multiple genres, the final sample is adjusted to 1,266 anime.

\subsection{Relational data}

We employ a graph in order to work with the network of users and anime. A graph \( G = (V, E) \) is a structure consisting of a set of vertices (nodes or actors) \( V \) and a set of edges (links or connections) \( E \), where the elements of \( E \) are pairs of the form \( e = \{u, v\} \), with \( u, v \in V \) \citep{kolaczyk2020}.
Within graph theory, there is a particular type known as a bipartite graph, also referred to as a two-mode network. This network consists of two distinct types of nodes, where connections only occur between nodes of different types. In this case, the sets of nodes are users and anime, respectively. Bipartite networks are fully investigated in \cite{guillaume2006}.

In order to analyze a graph, it is essential to determine its adjacency matrix. his matrix is a tabular representation where the rows and columns correspond to the individuals embedded in the system. In such a matrix, a value of one indicates the presence of a connection between two nodes, while a value of zero indicates its absence \citep{luke2015}. At this point an illustrative example is in order. Consider a network consisting of three users who watch four anime series: User 1 watches anime \(A\) and \(B\); User 2 watches anime \(B\) and \(C\); and User 3 watches anime \(A\), \(B\), \(C\), and \(D\). In this case, the adjacency matrix \(\mathbf{Y} = [y_{i,j}]\) is a \(3 \times 4\) matrix, in which rows represent users and columns represent anime series. Thus, \(y_{i,j} = 1\) indicates that user \(i\) watches anime \(j\), \(y_{i,j} = 0\) otherwise, for $i=1,2,3$ and $j=1,2,3,4$. The adjacency matrix in this toy example is given by:
\[
\mathbf{Y} = 
\begin{pmatrix}
1 & 1 & 0 & 0 \\
0 & 1 & 1 & 0 \\
1 & 1 & 1 & 1
\end{pmatrix}\,.
\]

In this way, consider an adjacency matrix \(\mathbf{Y}\) of size \(n \times p\) associated with a bipartite graph of users and anime. With this matrix, we can perform two insightful operations. First, \(\mathbf{Y}^\textsf{T}\mathbf{Y}\), which produces a \(p \times p\) matrix that reflects the number of users who simultaneously watch two specific anime series. Conversely, \(\mathbf{Y}\mathbf{Y}^\textsf{T}\), which produces an \(n \times n\) matrix that reflects the number of anime series watched by two specific users.
Regarding the toy example provided above, these matrices are:
\[
\mathbf{Y}^\textsf{T}\mathbf{Y} =
\begin{pmatrix}
0 & 2 & 1 & 1 \\
2 & 0 & 2 & 1 \\
1 & 2 & 0 & 1 \\
1 & 1 & 1 & 0
\end{pmatrix}
\qquad \text{and} \qquad
\mathbf{Y}\mathbf{Y}^\textsf{T} =
\begin{pmatrix}
0 & 1 & 2 \\
1 & 0 & 2 \\
2 & 2 & 0
\end{pmatrix}\,.
\]
These matrices reflect, respectively, the number of users watching pairs of anime and the number of anime watched in common by pairs of users.
It is important to note that in a graph, reflexive connections are not considered; therefore, the diagonal of the square matrices must be filled with structural zeros \citep{luke2015}.

Finally, given the specific nature of our analysis, we focus on the adjacency matrix regarding anime, denoted as \(\mathbf{Y}^\textsf{T}\mathbf{Y}\). Thus, we construct a new binary adjacency matrix where a value of 1 indicates that at least 75\% of users who watch either of the two anime series also watch both concurrently. This threshold, set higher than the typical 50\%, allows for the retention of more relevant information while reducing noise. By employing this criterion, we highlight only relevant connections, reflecting meaningful relationships among the anime series. The resulting binary adjacency matrix is essential for providing deeper insights into viewing patterns and user preferences within the dataset.

\subsection{Text data}

After establishing the graph that displays the significant connections among anime, we identify those variables associated with each node. These variables enable to enrich the analyzes of the underlying relational structures within the network \cite{luke2015}. In this case, each anime is described by a description, which can be analyzed using natural language processing (NLP). NLP integrates computational linguistics, which relies on rule-based systems to model human language, with statistical models and machine learning techniques. This combination allows computers and digital devices to recognize, understand, and generate both text and speech \citep{chowdhary2020}.

In this study, we implement bigram analysis as a natural language processing (NLP) tool by unifying all anime descriptions into a single corpus (large and structured set of texts that are collected for the purpose of linguistic analysis). Bigrams are tokens (single unit of text, which could be a word, a phrase or a symbol) composed of two consecutive words. In this way, we generate a graph where the vertices represent words, and the connections between them indicate that their consecutiveness is sufficiently important to establish a link. This process can be carried out in two ways: The first method involves removing stop words (such as articles, pronouns, and other non-informative words) before generating bigrams; the second method generates the bigrams first and then eliminates those containing stop words \citep{silge2017}. In this work, we select the methodology that yields the most word groupings, allowing for a clearer analysis of the content.

The connections considered sufficiently important, as mentioned above, are established as follows. We calculate the skewness of the bigrams' absolute frequencies, which measures the asymmetry of the distribution \citep{joanes_gill_1998}. The procedure involves systematically eliminating bigrams that occur fewer times than a given threshold. Then, using a skewness dispersogram plotted against thresholds, we select that threshold where the skewness no longer changes significantly or stabilizes. This method is similar in spirit to the elbow method used in Principal Component Analysis (PCA; \citealt{marutho2018elbow}).

Using the bigram network associated with the entire body of anime's description, we employ various community detection methods to cluster the corresponding description words \citep{kolaczyk2020}.
After implementing all available partition algorithms in \texttt{igraph} (excluding \texttt{optimal community structure}, which is specifically designed for small networks; \citealt{igraph_cluster_optimal}), we select the algorithm that maximizes modularity--a measure that quantifies the strength of a network's division into communities.

Graph clustering algorithms create a partition \( \mathcal{C} = \{\text{C}_1, \dots, \text{C}_K\} \) of the vertex set \( V \) of a graph \( G = (V, E) \). This partition is constructed such that the number of edges connecting vertices in \( \text{C}_k \) to vertices in \( \text{C}_\ell \) is relatively small compared to the number of edges connecting vertices within each \( \text{C}_k \) \citep{kolaczyk2020}. 
The modularity of a network with respect to a partition $\mathcal{C}$ measures the quality of the division or how separated the different types of vertices are from each other:
\[
\text{mod}(\mathcal{C}) = \frac{1}{2m} \sum_{i,j; i \neq j} \left( y_{i,j} - \frac{1}{2m} d_i d_j \right) \delta_{c_i, c_j}
\]
where
$\mathbf{Y}$ is the adjacency matrix, $m$ is the size of the graph (number of edges), $d_i$ is the degree of vertex $i$, $c_i$ is the membership group of vertex $i$, and $\delta_{x,y} = 1$ if $x = y$ and 0 otherwise.

Finally, we count the words in each description that belong to the identified clusters within the complete corpus.
Just to give you a toy example, suppose the bigram network has two clusters, namely, $\text{C}_1 = \{\text{day},\text{sun}\}$ and $\text{C}_2 = \{\text{adventure},\text{glory},\text{fantasy}\}$. 
If the description of a given anime is ``A young girl embarks on an epic adventure filled with glory until one day she achieves her goal'', this anime is attached to it two new variables. One variable assuming the value of 2 (for the words ``adventure'' and ``glory''), and the other assuming the value of 1 (for the word ``day'').
Once the new variables derived from the word clusters are created for each anime, we observe the influence of these variables when establishing a connection between two anime series using an Exponential Random Graph Model.

\subsection{Modeling}

To analyze how the count of words in the anime descriptions influences the anime network structure, we employ Exponential Random Graph Models (ERGMs; e.g., \citealt{frank1986markov}, \citealt{wasserman1996logit}, \citealt{robins2007introduction}). 
These models are specified analogously to generalized linear models (GLMs; e.g., \citealt{nelder1972}). They describe the conditional probability of a random adjacency matrix \( \mathbf{Y} \) for a simple, undirected binary network as follows \citep{hunter2008}:
\[
p(\boldsymbol{y} \mid \boldsymbol{\theta}) = \frac{1}{\kappa(\boldsymbol{\theta})} \exp\{\boldsymbol{\theta}^\textsf{T} g(\boldsymbol{y})\}\,,
\]
where
\( \boldsymbol{y} \) represents a realization of the adjacency matrix \( \mathbf{Y} \),
\( g(\boldsymbol{y}) = [g_1(\boldsymbol{y}), \ldots, g_K(\boldsymbol{y})]^\textsf{T} \) is a vector of statistics of \( \boldsymbol{y} \) and/or known functions of \( \boldsymbol{y} \) and node attributes \( \boldsymbol{x} \),
\( \boldsymbol{\theta} = [\theta_1, \ldots, \theta_K]^\textsf{T} \) is a vector of unknown parameters,
\( \kappa(\boldsymbol{\theta}) = \sum_{\boldsymbol{y}} \exp\{\boldsymbol{\theta}^\textsf{T} g(\boldsymbol{y})\} \) is the normalization constant.
The coefficients \( \boldsymbol{\theta} \) represent the magnitude and direction of the effects of \( g(\boldsymbol{y}) \) on the probability of observing the network.

Finally, to better understand the word groups in the entire corpus, we utilize artificial intelligence to assign meaning that allows for the identification of the most important topics addressed in the descriptions. Additionally, the aim is to expand the themes on which content should be written, making this feature more attractive when promoting an anime. Since certain nodes in any network are more important than others, artificial intelligence assigns importance from highest to lowest to optimally determine the main topic. This importance is calculated using eigenvector centrality, which indicates that a vertex is considered ``important'' if its neighbors are also ``central'' \citep{bonacich2007}:
\[
\text{c}_{\textsf{E}}(v) = \alpha\sum_{\{u,v\}\in E} c(u)\,
\]
where \( \boldsymbol{c} = (c_1, \ldots, c_n) \) is a solution to the eigenvector problem \( \mathbf{Y} \boldsymbol{c} = \alpha^{-1} \boldsymbol{c} \), where \( \mathbf{Y} \) is the adjacency matrix, \( \alpha^{-1} \) is the largest eigenvalue of \( \mathbf{Y} \), and \( \boldsymbol{c} \) is the corresponding eigenvector \citep{jalili2017}. The convention is to report the absolute values of the entries in \( \boldsymbol{c} \) \citep{newman2018}.

\section{Anime Case Study}

To initiate the analysis, we conduct an exploratory data analysis (EDA) on the corpus of all anime descriptions. A key step in this process is the removal of stop words, which enhances the clarity of the results. The most frequent words identified are presented in Table \ref{tab_word_freq}, with notable terms such as ``world,'' ``school,'' ``life,'' and ``girl.'' These words reflect prevalent themes and character archetypes in the anime genre, offering insights into the narratives and tropes that resonate with audiences.

\begin{table}[htb]
\centering
\begin{tabular}{cc}
\hline
Word & Frecuency \\ \hline
world  & 3147 \\ 
school & 2601 \\ 
life   & 2142 \\ 
ann    & 2036 \\ 
girl   & 1883 \\ 
story  & 1878 \\ 
series & 1836 \\ 
video  & 1770 \\ 
music  & 1711 \\ 
day    & 1608 \\ \hline
\end{tabular}
\caption{Frequency of the most common words in anime descriptions.}\label{tab_word_freq}
\end{table}

\begin{figure}[htb]
    \centering
    \includegraphics[width=0.5\linewidth]{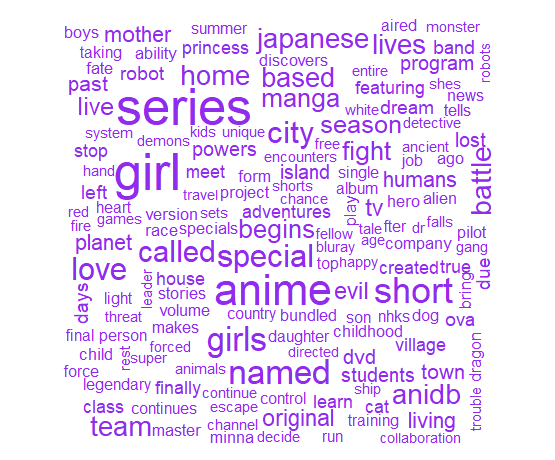}
    \caption{World cloud.}
    \label{fig_enter_label}
\end{figure}

The word cloud displayed in Figure \ref{fig_enter_label} highlights additional relevant words throughout the entire corpus of descriptions. This visualization allows for the rapid identification of common themes and elements present in anime descriptions. The most prominent words reflect a combination of school settings, personal relationships, fictional worlds, and elements of adventure and fantasy. By examining these terms, we gain a clearer understanding of the recurring motifs that define the genre and resonate with audiences.

\begin{figure}[htb]
    \centering
    \includegraphics[width=0.5\linewidth]{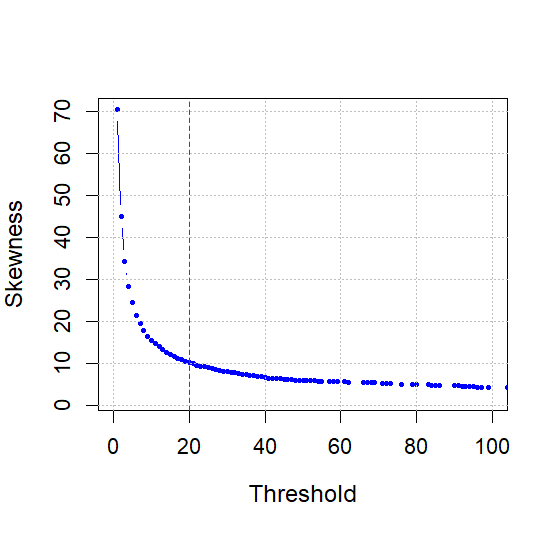}
    \caption{Skewness dispersogram plotted against the threshold.}
    \label{fig_umbral2}
\end{figure}

To construct the bigram network from the whole description corpus, we employ both approaches-—removing stop words before generating the bigrams and removing them afterward. 
The latter demonstrates better results in terms of meaningful connections; therefore, the remainder of the analysis is conducted using this method.
We establish a threshold of 20, as it appears that the skewness in the counts stabilizes at this value (see Figure \ref{fig_umbral2}).
Once the bigram network is established, we proceed to cluster the words by partitioning the network and identifying the respective groups. To accomplish this, we use all hierarchical clustering algorithms implemented in \texttt{igraph}, and we select the algorithm that maximizes modularity (see Table \ref{tab_mod}).
The results indicate that the Spinglass method is the clustering algorithm that maximizes modularity (generating 18 clusters) and, therefore, is selected for all subsequent procedures.

\begin{table}[htb]
\centering
\begin{tabular}{lc}
    \hline
    Algorithm            & Modularity \\ \hline
    Edge Betweenness     & 0.712      \\ 
    Fast Greedy          & 0.715      \\ 
    Infomap              & 0.687      \\ 
    Label Propagation    & 0.687      \\ 
    Leading Eigenvector  & 0.673      \\ 
    Leiden               & -0.010     \\ 
    Louvain              & 0.717      \\ 
    Spinglass            & 0.722      \\ 
    Walktrap             & 0.665      \\ \hline
\end{tabular}
\caption{Modularity of clustering in the bigram network.}\label{tab_mod}
\end{table}

Once the clusters are identified, we create 18 variables in the database, each representing the frequency of the words that comprise the group within the descriptions of each anime series.
These variables serve as the input in the modeling phase using an ERGM. Additionally, with the aid of artificial intelligence, the words in each group are analyzed to assign the cluster a meaningful interpretation. These topics are provided in Table \ref{tab_topics}. The Appendix provides a detailed analysis of the topic for each cluster, including key words and an exploration of the corresponding themes.


\begin{table}[htb]
\centering
\begin{tabular}{cl}
\hline
Cluster & Topic \\ \hline
1   & Sci-fi / Space \\ 
2   & Animation Techniques \\
3   & Character Design\\ 
4   & School Life\\
5   & Promotional Content\\
6   & Home Video Releases \\
7   &  Adventure \\
8   & School Positions \\
9   & Music / Band \\
10  & Romance \\
11  & Anime Adaptations \\
12  & Online Media \\ 
13  & Fantasy / Supernatural \\ 
14  & Future / Children's Content \\ 
15  & Character Focus \\ 
16  & Conflict / Resolution \\ 
17  & Friendship / Sports \\
18  & Historical / War\\ \hline
\end{tabular}
\caption{Topics of the word clusters.}
\label{tab_topics}
\end{table}

On the other hand, the bipartite network \(\mathbf{Y}\) of users and anime contains 11,730,932 links, as established after conducting the simple random sampling (SRS) described in the methodology section.
The projection of this bipartite network \(\mathbf{Y}^\textsf{T}\mathbf{Y}\) yields a weighted matrix of dimensions $1266 \times 1266$, where each entry represents the number of users who have viewed two specific anime series. As mentioned in the methodology, the network is binarized such that the entries of the new adjacency matrix take a value of 1 if at least 75\% of users who watch either of the two anime series also watch both concurrently, and assume a value of 0 otherwise. A comprehensive description of the topology of this network is presented in Table \ref{tab_metrics-summary}.

\begin{table}[htb]
    \centering
    \begin{tabular}{lc}
        \hline
        Statistic                & Value     \\ \hline
        Mean geodesic distance   & 1.8385    \\
        Mean degree              & 194.4780  \\
        SD degree                & 222.0349  \\
        Clique number            & 99        \\
        Density                  & 0.1615    \\
        Transitivity             & 0.3642    \\
        Associativity            & -0.5255   \\ \hline
    \end{tabular}
    \caption{Topology description for the anime network.}
    \label{tab_metrics-summary}
\end{table}

\begin{figure}[htb]
    \centering
    \includegraphics[width=0.6\linewidth]{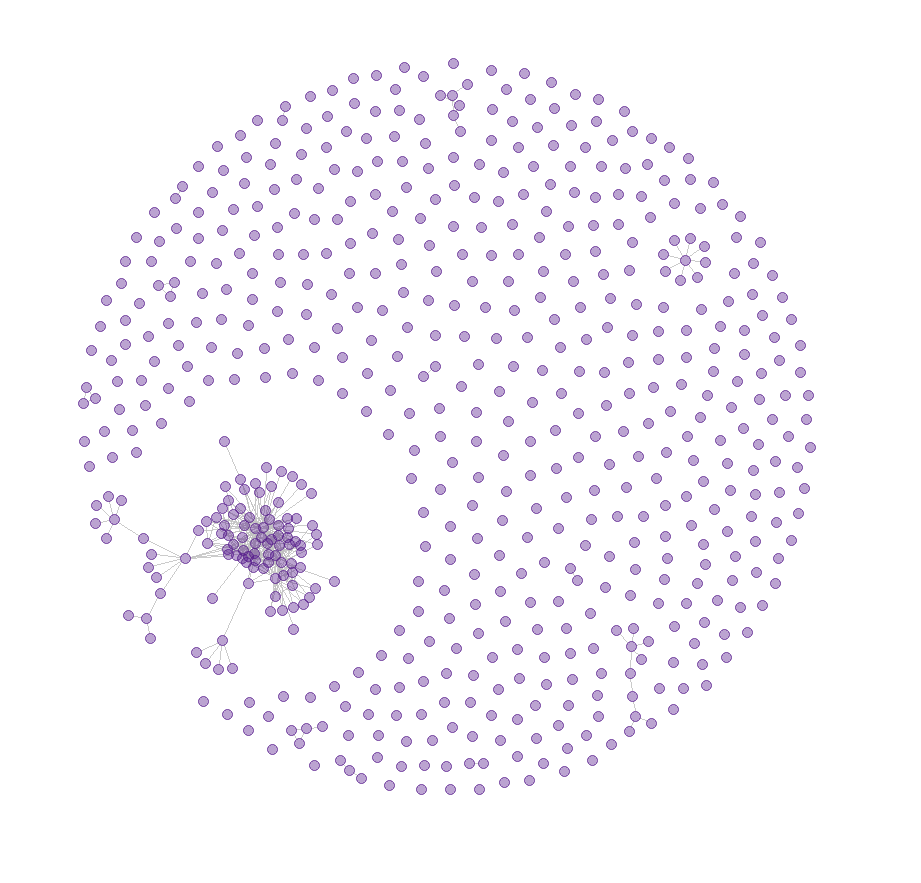}
        \caption{Anime network.}
    \label{fig_enter-label}
\end{figure}

Given that the original network is highly dense, as shown in Table \ref{tab_metrics-summary}, visual interpretation is challenging. To simplify the network and better reveal its structure, we employ the $k$-core method \citep{kong2019}. By applying this method and graphing only the nodes with a \( k \) value below the median, a clear trend emerges where nodes cluster around a central structure (see Figure \ref{fig_enter-label}). This central zone consists of the most important or influential anime within the network, serving as key points of connectivity and attraction for peripheral nodes. 

As shown in Table \ref{tab-maximum-likelihood-results}, fitting an ERGM to the anime network yielded significant results regarding the influence of specific word clusters on the formation of links. 
Our findings indicate that most clusters have significant coefficients, suggesting a clear association with the creation of connections within the network. 
Clusters 3, 5, 6, 9, 12, and 14 exhibit positive coefficients, indicating that these word groups actively promote link formation within the anime network. Among these, Cluster 14 is particularly notable, as it possesses a high coefficient of 0.1725. This suggests that the themes encapsulated by this cluster—potentially reflecting popular tropes or compelling narratives—are strongly correlated with users' propensity to watch multiple anime concurrently. The presence of such influential word groups may enhance viewer engagement and foster interconnectedness, thereby shaping the overall structure of the network.

Conversely, Clusters 1 and 18 display significant negative coefficients, indicating a considerable inhibitory effect on link formation within the network. Notably, Cluster 18 stands out with a coefficient of -0.3159, suggesting that the themes represented in this cluster may deter users from engaging with multiple anime simultaneously. This could imply that certain narrative elements or contexts present in the descriptions associated with this cluster are less appealing or incompatible, leading to reduced viewer overlap. While clusters 2 and 16 do not reach statistical significance, the overall analysis underscores that the majority of clusters exert a clear influence on the network's connectivity. This pattern highlights the complex interplay of word groups and their thematic implications in shaping viewer behaviors and preferences, reinforcing the idea that not all content resonates equally within the anime community.

Notably, these findings align with the broader narrative that anime descriptions significantly shape audience behavior, suggesting that marketing strategies could benefit from emphasizing the most influential word clusters to enhance viewer engagement. Additionally, in the same spirit of \cite{gelman2013bayesian}, the goodness-of-fit of the model (not shown here) shows that the relationships captured by the ERGM reflect genuine social patterns among viewers, reinforcing the notion that language and thematic representation are critical components in understanding the anime viewing experience. This understanding can inform content creators and marketers alike, guiding them in crafting compelling descriptions that resonate with target audiences and foster community among anime enthusiasts.

\begin{table}[htb]
    \centering
    \begin{tabular}{lrrrr}
    \hline
    \textbf{Coefficient} & \textbf{Estimate} & \textbf{Standard Error} & \textbf{$z$-value} & \textbf{$\Pr(>|z|)$} \\
    \hline
    edges              & -1.914911 & 0.010303 & -185.858 & $< 1 \times 10^{-4}$ \\ 
    cluster 1          & -0.171144 & 0.004292 &  -39.877 & $< 1 \times 10^{-4}$ \\ 
    cluster 2          & -0.005335 & 0.007888 &   -0.676 & 0.499 \\ 
    cluster 3          &  0.117761 & 0.007195 &   16.367 & $< 1 \times 10^{-4}$ \\ 
    cluster 4          &  0.012500 & 0.002489 &    5.022 & $< 1 \times 10^{-4}$ \\ 
    cluster 5          &  0.117565 & 0.003165 &   37.143 & $< 1 \times 10^{-4}$ \\ 
    cluster 6          &  0.073631 & 0.002998 &   24.560 & $< 1 \times 10^{-4}$ \\ 
    cluster 7          &  0.046664 & 0.006809 &    6.853 & $< 1 \times 10^{-4}$ \\ 
    cluster 8          &  0.049898 & 0.006691 &    7.458 & $< 1 \times 10^{-4}$ \\ 
    cluster 9          &  0.148945 & 0.004522 &   32.939 & $< 1 \times 10^{-4}$ \\ 
    cluster 10         & -0.016673 & 0.004999 &   -3.335 & 0.001 \\ 
    cluster 11         &  0.022819 & 0.002202 &   10.364 & $< 1 \times 10^{-4}$ \\ 
    cluster 12         &  0.068163 & 0.004575 &   14.899 & $< 1 \times 10^{-4}$ \\ 
    cluster 13         &  0.012382 & 0.002831 &    4.374 & $< 1 \times 10^{-4}$ \\ 
    cluster 14         &  0.172535 & 0.005702 &   30.258 & $< 1 \times 10^{-4}$ \\ 
    cluster 15         & -0.025920 & 0.002562 &  -10.119 & $< 1 \times 10^{-4}$ \\ 
    cluster 16         & -0.007162 & 0.009094 &   -0.788 & 0.431 \\ 
    cluster 17         &  0.031619 & 0.002956 &   10.696 & $< 1 \times 10^{-4}$ \\ 
    cluster 18         & -0.315912 & 0.008485 &  -37.231 & $< 1 \times 10^{-4}$ \\ \hline
    \end{tabular}
   \caption{ERGM results in the anime network.}
    \label{tab-maximum-likelihood-results}
\end{table}

\section{Discussion}

This study demonstrates that the analysis of bipartite networks, in conjunction with natural language processing techniques, is an effective tool for understanding how textual descriptions influence the formation of anime or any other user communities. The use of Exponential Random Graph Models (ERGMs) enable the identification of significant clusters of words on the creation of links between anime titles.

Thus, we conclude that:
\begin{itemize}
    \item \textbf{Positive influence of certain clusters:} Word groups such as Cluster 3 (Character Design), Cluster 5 (Promotional and Educational Content), Cluster 6 (Video Format Releases), Cluster 7 (Adventure and Travel), Cluster 8 (Student Life and Council Roles), Cluster 9 (Music and Bands), Cluster 11 (Anime Series and Adaptations), Cluster 13 (Fantasy and Supernatural), and Cluster 14 (Future and Children’s Content) exhibited a significant positive influence on link formation. This suggests that utilizing words associated with these themes increases the likelihood of connections between different anime titles.

    \item \textbf{Negative influence of other clusters:} In contrast, Cluster 1 (Science Fiction and Defense of Humanity), Cluster 10 (Romance and Relationships), Cluster 17 (Friendship and Sports), and Cluster 18 (Wars and Historical Conflicts) showed significant negative coefficients. This indicates that a higher frequency of terms related to these clusters decreases the probability of link formation between anime.

    \item \textbf{Clusters without significant influence:} Clusters such as Cluster 2 (Animation Techniques) and Cluster 16 (Final Battles and Conflict Resolution) did not demonstrate significant influence on link formation, as their $p$-values exceeded 0.05.

    \item \textbf{Importance of detailed descriptions:} Including well-structured descriptions on anime platforms may enhance community formation among users, thereby improving overall user experience and facilitating more accurate recommendations.
\end{itemize}

Based on the study's findings, we strongly recommend:
\begin{itemize}
    \item \textbf{Optimization of descriptions:} Anime platforms should optimize their descriptions by incorporating terms and phrases that have been shown to positively influence link formation. This could strengthen community cohesion and increase user interaction.

    \item \textbf{Improvement of recommendation systems:} Integrating cluster analysis of words into recommendation systems could enhance their accuracy, suggesting anime that are not only popular but also relevant based on textual descriptions.

    \item \textbf{Future research directions:} It is advisable to conduct additional studies that expand this analysis to a larger dataset and other entertainment genres, to validate the applicability of this methodology across different contexts.

    \item \textbf{Clarification of cluster themes:} Establishing a clear theme for the word clusters may be challenging due to the apparent lack of correlation among some words.
\end{itemize}

In conclusion, this study underscores the importance of textual content in forming user networks and suggests that a combined approach of text analysis and graph theory can provide valuable insights to enhance user experience on anime platforms and beyond.

\section*{Statements and Declarations}

The authors declare that they have no known competing financial interests or personal relationships that could have appeared to influence the work reported in this article.

\bibliography{bibliografia.bib}
\bibliographystyle{apalike}

\appendix

\section{Detailed Topic Analysis by Clusters}
\label{sec:analisis_clusters}

\begin{itemize}
    \item \textbf{Cluster 1: Science Fiction / Space}
    \begin{itemize}
        \item \textbf{Keywords}: home, save, planet, return, returns, earth, humanity, federation, protect
        \item \textbf{Theme}: This cluster is related to the defense of Earth and humanity in science fiction and space adventure settings.
    \end{itemize}

    \item \textbf{Cluster 2: Animation Techniques}
    \begin{itemize}
        \item \textbf{Keywords}: animation, motion, stopmotion, toei, puppet, stop
        \item \textbf{Theme}: Focuses on animation and production techniques, including stop-motion and work from studios like Toei.
    \end{itemize}

    \item \textbf{Cluster 3: Character Design}
    \begin{itemize}
        \item \textbf{Keywords}: main, characters, character, featuring, designs
        \item \textbf{Theme}: This cluster groups information related to the design and development of main characters in anime.
    \end{itemize}

    \item \textbf{Cluster 4: School and Everyday Life}
    \begin{itemize}
        \item \textbf{Keywords}: school, life, day, girls, lives, ordinary, normal, boys, junior, students, idol, middle, boarding, elementary, real, daily, everyday, death, peeping, peaceful, change, academy, cinderella, private
        \item \textbf{Theme}: Focused on school life, the daily experiences of students, and the challenges they face in an educational environment.
    \end{itemize}

    \item \textbf{Cluster 5: Promotional and Educational Content}
    \begin{itemize}
        \item \textbf{Keywords}: video, safety, promotional, features, nhks, game, program, games, directed, traffic, fire, affic, minna, featured, uta, mobile, card, suit, gundam
        \item \textbf{Theme}: Content centered on promotional videos, educational programs, and games, particularly for franchises like Gundam.
    \end{itemize}

    \item \textbf{Cluster 6: Video Format Releases}
    \begin{itemize}
        \item \textbf{Keywords}: episode, episodes, special, tv, included, dvd, bundled, aired, specials, volume, bluray, ova, cap, unaired, released, bluraydvd, release, volumes, version, bonus, bddvd, releases, limited, album, edition
        \item \textbf{Theme}: Focuses on the release of special episodes, DVD and Blu-ray editions, and additional content.
    \end{itemize}

    \item \textbf{Cluster 7: Adventure and Journeys}
    \begin{itemize}
        \item \textbf{Keywords}: begins, journey, sets, embark, embarks
        \item \textbf{Theme}: This cluster deals with adventures and quests where characters embark on significant or heroic journeys.
    \end{itemize}

    \item \textbf{Cluster 8: Student Life and Roles}
    \begin{itemize}
        \item \textbf{Keywords}: student, council, college, university, transfer, president
        \item \textbf{Theme}: Focuses on student life and roles in student councils, as well as transfers between institutions.
    \end{itemize}

    \item \textbf{Cluster 9: Music and Bands}
    \begin{itemize}
        \item \textbf{Keywords}: song, irodorimidori, franchise, theme, band, focuses, rock, fictional, japanese, traditional
        \item \textbf{Theme}: This cluster focuses on music, bands, and traditional Japanese elements, often within a fictional context.
    \end{itemize}

    \item \textbf{Cluster 10: Romance and Relationships}
    \begin{itemize}
        \item \textbf{Keywords}: love, live, true, fall, falls, humans, action, identity, nature
        \item \textbf{Theme}: Romantic relationships and the emotional development of characters in both personal and action-packed situations.
    \end{itemize}

    \item \textbf{Cluster 11: Anime Series and Adaptations}
    \begin{itemize}
        \item \textbf{Keywords}: series, anime, short, story, based, original, animated, manga, shorts, television, film, ducational, added, encyclopedia, adaptation, movie, stories, films, tells, takes, revolves, continues, centers, kouji, artist, feature, festival, synopsis, nanke
        \item \textbf{Theme}: Adaptations of stories in various formats, from short series to feature films and anime movies.
    \end{itemize}

    \item \textbf{Cluster 12: Online Media Platforms}
    \begin{itemize}
        \item \textbf{Keywords}: music, official, al, videos, ian, youtube, twitter, posted, site, website, channel
        \item \textbf{Theme}: Use of online platforms like YouTube and Twitter for promoting anime-related content.
    \end{itemize}

    \item \textbf{Cluster 13: Fantasy and the Supernatural}
    \begin{itemize}
        \item \textbf{Keywords}: powers, magical, world, supernatural, human, fantasy, fate, magic, people, digital, conquer, parallel, entire, domination, travels, dream, spirit, peace, race, alien
        \item \textbf{Theme}: Stories exploring fantastical worlds, magical powers, and supernatural elements.
    \end{itemize}

    \item \textbf{Cluster 14: Future and Children's Content}
    \begin{itemize}
        \item \textbf{Keywords}: set, childrens, box, future, book, distant, picture, drama
        \item \textbf{Theme}: This cluster focuses on content set in the future and stories aimed at a younger audience.
    \end{itemize}

    \item \textbf{Cluster 15: Character Development}
    \begin{itemize}
        \item \textbf{Keywords}: girl, boy, named, yearold, mysterious, beautiful, called, meets, strange, cute, woman, organization, power, eye, secret
        \item \textbf{Theme}: Stories centered on character development, their relationships, and personal mysteries.
    \end{itemize}

    \item \textbf{Cluster 16: Conflicts and Resolutions}
    \begin{itemize}
        \item \textbf{Keywords}: final, battle
        \item \textbf{Theme}: Focuses on final battles and the resolution of significant conflicts.
    \end{itemize}

    \item \textbf{Cluster 17: Friendship and Sports}
    \begin{itemize}
        \item \textbf{Keywords}: friends, days, baseball, time, spends, family, childhood, makes, close, team, space, machine, spend, ago, friend, nohara, soccer, basketball, station, outer
        \item \textbf{Theme}: Friendships and the importance of sports in the characters' daily lives.
    \end{itemize}

    \item \textbf{Cluster 18: Historical Contexts and Wars}
    \begin{itemize}
        \item \textbf{Keywords}: click, update, war, information, grail, civil, ii, holy
        \item \textbf{Theme}: Refers to historical contexts and wars, especially in tales of epic battles and major conflicts.
    \end{itemize}
\end{itemize}

\end{document}